\documentstyle[11pt,newpasp,twoside,natbib,psfig]{article}
\bibpunct[; ]{(}{)}{,}{a}{}{,}

\newcommand\phn{\phantom{0}}

\begin{document}

\title{Improved Bounds on Violation of the Strong Equivalence Principle}

\author{Z. Arzoumanian}
\affil{Universities Space Research Association, Laboratory for
High-Energy Astrophysics, NASA-GSFC, Greenbelt, MD~20771}

\begin{abstract}
I describe a unique, 20-year-long 
timing program for the binary pulsar B0655+64, the
stalwart control experiment for measurements of gravitational
radiation damping in relativistic neutron-star binaries. Observed
limits on evolution of the B0655+64 orbit provide new bounds on 
the existence of dipolar gravitational radiation, and hence on
violation of the Strong Equivalence Principle.
\end{abstract}

\section{Introduction}

%\looseness-1
PSR B0655+64, in a highly circular one-day orbit with a $\sim 0.8$
$M_{\sun}$ white-dwarf companion, serves as a control experiment for
measurements of orbital decay in the highly relativistic
double-neutron-star binaries: General Relativity (GR) has predicted
equally well the strong back-reaction to gravitational radiation for
PSRs B1913+16 \citep[][see also Weisberg \& Taylor, this
volume]{tay93b} and B1534+12
%\citep{2002astro.ph..8357S,1998ApJ...505..352S}, and 
\citep{1998ApJ...505..352S}, and 
the absence of detectable orbital evolution for PSR B0655+64 over
two decades. The long-term stability of the B0655+64
orbit sets unique bounds on departures from GR that give rise
to dipolar gravitational radiation \citep{arz95,gol92}, the
existence of which would represent a violation of the Strong
Equivalence Principle (SEP), one of the basic tenets of GR.
\citet{2002PhRvD..66b4040G} examine the theoretical basis for
dipolar gravitational radiation and suggest that bounds from binary
pulsars may be competitive with future satellite experiments
dedicated to probing SEP violation.  

A definitive analysis of the available data for PSR B0655+64, and
implications for a variety of alternative theories of gravitation,
will be published separately (Arzoumanian et al.\ 2003, in
preparation). Following recent observations to extend our long-term
monitoring program, I present here preliminary results on the
orbital evolution of PSR B0655+64.

\subsection{Binary Pulsars and the Strong Equivalence Principle}

%Relativistic theories of gravity can be tested with pulsar binaries
%in two ways, by establishing---or refuting---the validity of
%fundamental assumptions, and by demanding the self-consistency of
%predicted ``post-Newtonian'' effects (such as precession of
%periastron) in any one theory. The results of individual tests, even
%from different pulsars, can be combined to provide strong
%constraints on entire families of alternative theories
%\citep[e.g.,][]{twdw92}.

%Violation of the Strong Equivalence Principle implies that the
%response of a body to an external gravitational field depends on the
%gravitational make-up of the body. Two objects with different
%gravitational binding energies, for example, will fall at different
%rates.  
The Strong Equivalence Principle posits that the response of a body
to an external gravitational field is independent of the
gravitational self-energy of the body; it would be violated if, for
example, two objects with different gravitational binding energies
were observed to fall at different rates.
SEP is satisfied by postulate within GR; as a consequence,
the lowest allowed multipole order of gravitational radiation is the
``electric'' quadrupole.  If SEP does not hold, however, emission of
dipolar gravitational radiation is allowed, and lends itself in
principle to indirect detection through decay of a binary orbit
at a rate inconsistent with the quadrupole-order prediction of GR.
Because self-gravitational effects are important in neutron stars,
SEP can be tested with pulsar systems. 
%in a regime inaccessible to solar-system experiments.  

The most thoroughly studied theory of
gravitation that violates SEP, the so-called ``Brans-Dicke''
scalar-tensor theory \citep{bd61}, invokes the existence of a
scalar gravitational field that couples to matter through a single
%that acts as a spatially varying gravitational constant.  
%The scalar field in the Brans-Dicke theory enters into the field
%equations that determine the structure of space-time but does not
%act directly on the motion of bodies. The importance, in the field
%equations of gravitation, of the energy-momentum tensor due to the
%scalar field 
%Coupling of the scalar field to matter is governed by a single
dimensionless parameter, $\omega_{B\!D}$. The Brans-Dicke theory
becomes indistinguishable from GR as $\omega_{B\!D}$ becomes
arbitrarily large; solar-system experiments currently place a lower
bound $\omega_{B\!D} \ga 500$. Other alternatives to GR are also
constrained by limits on dipolar gravitational radiation \citep[see,
e.g.,][]{de92}: recent observations of distant type Ia supernovae
suggesting that the expansion of the universe is accelerating 
%(Reiss et al.\ 1998, AJ 116, 1009; Perlmutter et al.\ 1999, ApJ 517, 565)
have prompted renewed interest in theories involving scalar fields
(e.g., ``quintessence,'' \citealp{1998Ap&SS.261..303C}) or
long-range forces
%(Caldwell et al.\ 1998, Ap\&SS 261, 303), 
that may be responsible for the acceleration.  Such fields
would radiate predominantly at dipole order. 
%\citet{fie56}, \citet{jor59}, and 

The various contributions to observed changes in the orbital period
$P_b$ of a binary system (neglecting tidal and mass-transfer
effects) can be expressed as %without theoretical bias by 
\begin{equation}
\label{brkdown.eq}
\left(\frac{\dot P_b}{P_b}\right)_{\rm obs} = \left(\frac{\dot
P_b}{P_b}\right)_Q + \left(\frac{\dot P_b}{P_b}\right)_D +
\left(\frac{\dot P_b}{P_b}\right)_{\dot G} + \left(\frac{\dot
P_b}{P_b}\right)_{\bf a},
\end{equation}
where the subscripts $Q$, $D$, $\dot G$, and $\bf a$ denote the
effects of quadrupolar and dipolar gravitational radiation, a change in
the gravitational constant with time, and the varying Doppler shift from a
relative acceleration of the solar system and pulsar
binary\footnote{To correct for Galactic accelerations and the
``Shklovskii'' effect, I follow 
\citet{dt91}, using a new proper-motion measurement for B0655+64: 
$\mu = 6.8\pm1.1\;{\rm mas\,yr}^{-1}$.}.
%In GR, the dipolar radiation and $\dot G$ terms are exactly zero,
%while the quadrupole term is given, for a circular orbit, by
%\citep[e.g.,][]{ear75}
%\begin{equation}
%\label{pbdot.eq}
%\dot P_b = -\frac{192\pi}{5}n^{5/3} m_1m_2M^{-1/3}\,T_{\sun}^{5/3},
%\end{equation}
In a large class of
metric theories of gravitation, the 
radiation terms for a circular orbit are given by
\citep{ear75,wil81,gol92}
\begin{eqnarray}
\label{quad.eq}
\left(\frac{\dot P_b}{P_b}\right)_Q &=& -\frac{96}{5}{\cal
G}^{-4/3}\left(\frac{\kappa_1}{12}\right)
n^{8/3}m_1m_2M^{-1/3}T_{\sun}^{5/3} \\
\label{dip.eq}
\left(\frac{\dot P_b}{P_b}\right)_D &=& -\kappa_D{\cal
G}^{-2}(s_1-s_2)^2n^2\frac{m_1m_2}{M}T_{\sun},
\end{eqnarray}
where $n \equiv 2\pi/P_b$, $m_1$ and $m_2$ are the pulsar and
companion masses in solar units, $M \equiv m_1 + m_2$, and $T_{\sun}$
is the mass of the Sun in units of time. The dimensionless
parameters $\kappa_1$, $\kappa_D$, and $\cal G$ describe the
strength of quadrupolar and dipolar radiation and the effective
gravitational constant in a given theory.
%in the vicinity of the system. 
In GR, $\kappa_1=12$,
$\kappa_D=0$, and ${\cal G} = 1$;
%so that Eqs.~\ref{quad.eq} and \ref{dip.eq} reduce to Eq.~\ref{pbdot.eq}.  
in the Brans-Dicke theory, 
$\kappa_D{\cal G}^{-2} = 2(2+\omega_{B\!D})^{-1}$.
At the level of the current constraint on the dipole term $D$,
the $\dot G$ contribution to $\dot P_b$ in Eq.~\ref{brkdown.eq} can
be neglected \citep{arz95,ktr94}.

The quantities $s_1$ and $s_2$ in Eq.~\ref{dip.eq} represent the
``sensitivities'' of the orbiting objects, the fractional change in
binding energy of each star with $G$,
\begin{equation} 
\label{sens.eq}
s = -\left(\frac{\partial \ln m}{\partial \ln G}\right)_N,
%= \frac{3}{2}\left [1-\left(\frac{\partial \ln
%m}{\partial \ln N}\right)_G \right],
\end{equation}
where $N$ is the total number of baryons. The sensitivity of white
dwarfs is negligible;
neutron stars are thought to have sensitivities in the range
$0.15 < s < 0.40$ \citep{wz89}, with larger values corresponding to 
``softer'' equations of state. Results of X-ray burst oscillation
modeling \citep[e.g.,][]{2002ApJ...564..353N}, and the recent
detection of atmospheric absorption lines from a neutron star
\citep{cottam02}, suggest that fairly stiff equations of state are
appropriate. I therefore adopt a fiducial value $s = 0.2$ below. It
is worth noting that the difference $(s_1 - s_2)^2$ in
Eq.~\ref{dip.eq} strongly suppresses dipolar radiation from NS-NS
binaries like B1913+16, so that the celebrated agreement of the
latter's orbital decay rate with the GR prediction does not usefully
constrain the existence of gravitational radiation at dipole order.
In scalar-tensor theories, gravitational binding energy takes
on the role of an effective gravitational ``charge.'' Neutron
star-white dwarf (NS-WD) binaries can therefore be expected to
copiously emit dipolar radiation, if such radiation
exists, because of the large difference in gravitational self-energy
between the component stars.  
%Furthermore, orbital damping due to a
%dipolar radiation term can be much more important than the
%quadrupolar contribution because it typically depends on a smaller
%power of $v_{\rm orb}/c$ \citep{wz89}.  
%As shown below, our ability to constrain the existence of dipolar
%radiation is limited, in practice, by our insufficient understanding
%of NS equations of state and, to a lesser extent, by our inability
%to predict the amount of quadrupolar radiation in a given system
%with unknown component masses.

\subsection{PSR B0655+64}
\label{0655.sec}

%\looseness-1
PSR B0655+64 was discovered during a survey of the Northern sky made
with the NRAO 300 Foot telescope \citep{dbtb82}.  Timing observations
began soon thereafter, and the companion was optically identified as
a massive white dwarf by \citet{kul86}.
%The very small spin-down rate of this pulsar,
%$\dot P = 7\times 10^{-19}$, suggests that it is a recycled neutron
%star with a relatively weak magnetic field.  The extremely circular
%orbit, $e < 3\times 10^{-5}$, implies tidal circularization occured
%before the companion's progenitor collapsed to degeneracy, but it is
%unclear whether the neutron star was formed through accretion-induced
%collapse of a white dwarf \cite{mic87,bg90} or was spun-up by
%accretion after being formed in a supernova explosion. 
%Because of the high companion mass, this system has some features in
%common with the double neutron-star binaries B1913+16 and B1534+12;
%it seems likely that the companion narrowly
%escaped becoming a neutron star in a supernova of its own.
%\scite{bro94} discusses the possible evolution of the system from a
%pair of Wolf-Rayet (helium) stars.
If we assume pulsar and companion masses of 1.4 $M_{\sun}$ and 0.8 $M_{\sun}$,
Eq.~\ref{quad.eq} predicts a rate of orbital decay within GR,
%\begin{equation}
\(
\dot P_b^{\rm GR} = -2\times 10^{-14},
\)
%\end{equation}
corresponding to a decay timescale for the orbit of $\tau_Q \sim
150$~Gyr.

%The mass function of the PSR B0655+64 system implies a minimum
%companion mass of 0.67\,\Msun.  In the absence of other measurable
%quantities that can be linked to the component masses and
%inclination, no further characterization of the system can take
%place. However, the small orbital separation and large velocity
%($v^{\rm orb}/c = 3\times 10^{-4}$) suggest that relativistic
%effects may be important. Since the orbital eccentricity is very
%small, precession of periastron is unmeasurable. For the same
%reason, the gravitational redshift/time dilation parameter $\gamma$
%is degenerate with the Keplerian orbital parameters and cannot be
%measured. 

\section{Observations and Results}

The current data span 20 years, including observations made with the
NRAO Green Bank 300 Foot \citep{td88}, 140 Foot \citep{btd82,antt94},
and GBT telescopes, as well as the NRAO VLA \citep{tho91b} and
Jodrell Bank Lovell \citep{jl88} telescopes.  We carried out
intensive observing campaigns, to obtain good coverage of all
orbital phases at a single epoch, every few years with the 140 Foot
and recently with the GBT, and these data provide the interesting
constraints on orbital evolution.

Pulse times-of-arrival (TOAs) were derived from observations following
standard procedures and analyzed with the {\sc Tempo} software
package,
%\citep{tw89}, 
using %the JPL {\sc de200} planetary ephemeris \citep{sta93} and 
the {\sc ell1} binary timing formula
\citep{2001MNRAS.326..274L}. A least-squares fit of the entire dataset
for rotational, astrometric, and orbital parameters produces
residuals consistent with statistical fluctuations. 
%and reveals no long-term trend requiring
%a spin period second-derivative, $\ddot P$, in the timing model,
%demonstrating that B0655+64 is an extremely stable rotator and a
%useful clock for relativistic studies.

%Although still far from detecting period changes at the
%level of the GR prediction for quadrupolar radiation, the current
%dataset places 

\begin{figure}[h]
\centerline{\psfig{file=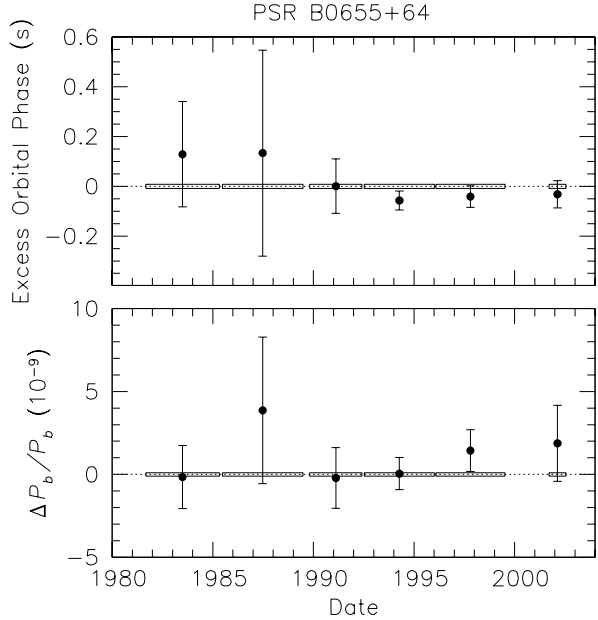,width=3.25in}}
\caption{\label{orbevol.fig}Orbital evolution of PSR B0655+64. 
%The top panel shows observed minus computed orbital phase shifts as
% measured from the epoch of ascending node; the lower panel
% displays fractional orbital period changes with time. 
Box outlines depict ranges of dates
over which phase and period measurements were made.}
\end{figure}

Orbital phase residual and period measurements over time are shown
in Figure~\ref{orbevol.fig}.  The constancy of $P_b$ confirms the
result of the global timing solution:  the available data bound
changes in orbital period at the level
%\begin{equation} 
\(
|\dot P_b|_{\rm obs} < 1.5\times 10^{-13}\,\,\,(1\sigma), 
\)
%\end{equation} 
a limit $\sim 7$ times the GR prediction. 
%While the effects of
%GR cannot yet be observed in this system, the available constraint
%on orbital decay is nonetheless interesting.  
Accounting for a
small correction for relative acceleration (Eq.~\ref{brkdown.eq}),
we then have a $2\sigma$ upper limit on orbital evolution, 
\begin{equation} 
\label{pbdlim.eq} 
\left|\frac{\dot P_b}{P_b}\right|_{\rm obs} < 1.0\times 10^{-10}\,{\rm yr}^{-1}
\,\,\,(2\sigma).
\end{equation}

\begin{figure}[h]
\centerline{\psfig{file=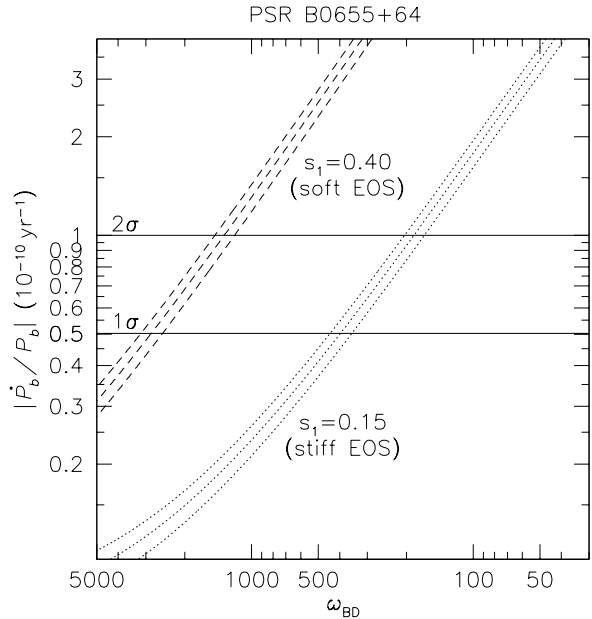,width=3.25in}}
\caption{\label{bdomega.fig}Bounds on %the Brans-Dicke parameter
$\omega_{B\!D}$ from PSR B0655+64 (after \citealp{wz89}).}
\end{figure}

The resulting constraints on $\omega_{B\!D}$
are summarized in Fig.~\ref{bdomega.fig}. Horizontal lines
depict the current $1\sigma$ and $2\sigma$ limits on orbital decay in
the PSR B0655+64 system.  The curving dashed and dotted lines are the
expected orbital period changes due to emission of gravitational
radiation through quadrupole order,
Eqs.~\ref{quad.eq}--\ref{dip.eq}, for NS sensitivities
from soft and stiff equations of state respectively.  The three
curves in each set represent the ($m_1$, $m_2$) pairs  (1.30,
0.7), (1.35, 0.8), and (1.40, 0.9), so that total system mass
increases to the upper left.
Constraints on $\omega_{B\!D}$ then lie at the intersections of the
dashed or dotted curves with the measured upper limits on $\dot P_b$.
From Eqs.~\ref{quad.eq}, \ref{dip.eq}, and \ref{pbdlim.eq}, we have
in general
\begin{equation}
\label{kd.eq}
\kappa_D{\cal G}^{-2} < 0.006\left(\frac{s}{0.2}\right)^{-2}\,\,\,(2\sigma),
\end{equation}
and for the Brans-Dicke theory specifically,
%\begin{equation}
%\omega_{B\!D} \ga 320\left(\frac{s}{0.2}\right)^2\,\,\,(2\sigma).
%\end{equation}
\(
\omega_{B\!D} \ga 320\,(s/0.2)^2\,\,\,(2\sigma).
\)

\section{Discussion}

%Pulsar binaries in which the companion star is a white dwarf (NS-WD
%systems) are uniquely suited to test for the existence of dipolar
%radiation of a scalar gravitational field.  Neither the double
%neutron-star binaries (e.g., B1913+16) nor wide-orbit millisecond
%pulsar binaries like B1855+09 \citep{ktr94} 
%%(Kaspi, Taylor \& Ryba %1994, ApJ 428, 713) 
%will emit copiously at dipole order even if such radiation exists:
%the former because the component stellar masses and internal
%properties are nearly equal, the latter because it is far less
%relativistic than B0655+64, with orbital period an order of
%magnitude longer.

\begin{table}
\begin{center}
\caption{\label{nswd.tab} Relativistic NS-WD systems. Spin and orbital
periods are shown alongside spin-down luminosity
($\dot E$), orbital separation ($A$) and timescales for orbital
evolution from quadrupolar ($\tau_Q$) and dipolar ($\tau_D$)
radiation. The potential for heating and tidal interactions between
the component stars increases rapidly for increasing $\dot E$ and
decreasing $A$.}
\begin{tabular}{lcccccc}
\tableline
\multicolumn{1}{c}{{PSR}} & {$P$} & {$P_b$} &
	{\rule{0cm}{4mm}$\log(\dot E$[erg/s])} & {$A$} & {$\tau_Q$} &
	{$\tau_D$} \\
	& {(ms)} & {(d)}	& {} &
	{(lt-s)} & {(Gyr)} & {(Myr)} \\
\tableline
J1141$-$6545 & 393.9 & 0.20 & {33.4} & {\phn4.5} & \phn\phn2 & \phn2 \\
J0751+1807 & \phn\phn3.5 & 0.26 & {33.8} & {\phn4.1} & \phn20 & 15 \\
J1757$-$5322 & \phn\phn8.9 & 0.45 & {33.2} & {\phn6.4} & \phn20 & 15 \\
{B0655+64} & {195.6} & {1.03} & {30.6} & {11.3} & 150 & 60 \\
J1012+5307 & \phn\phn5.3 & 0.60 & {33.6} & {\phn6.8} & 200 & 90 \\
J1435$-$6100 & \phn\phn9.3 & 1.35 & 33.1 & 14.2 & 250 & 90 \\
%J1157$-$5112 & \phn43.6 & 3.51 & 31.8 & 27.9 & 2400 & 515 \\
\tableline
\tableline
\end{tabular}
\end{center}
\end{table}
%J1141$-$6545 & 393.9 & 0.20 & {33.4} & {\phn4.5} & \phn\phn\phn2 & \phn\phn2 \\
%J0751+1807 & \phn\phn3.5 & 0.26 & {33.8} & {\phn4.1} & \phn\phn20 & \phn15 \\
%J1757$-$5322 & \phn\phn8.9 & 0.45 & {33.2} & {\phn6.4} & \phn\phn20 & \phn15 \\
%{B0655+64} & {195.6} & {1.03} & {30.6} & {11.3} & {\phn150} & {\phn60} \\
%J1012+5307 & \phn\phn5.3 & 0.60 & {33.6} & {\phn6.8} & \phn200 & \phn90 \\
%J1435$-$6100 & \phn\phn9.3 & 1.35 & 33.1 & 14.2 & \phn250 & \phn90 \\
%J1157$-$5112 & \phn43.6 & 3.51 & 31.8 & 27.9 & 2400 & 515 \\

Prospects for improving bounds on SEP violation and the
existence of dipolar gravitational radiation are good.
\citet{dt92} show that, for constant data quality, measurement
uncertainty for $\dot P_b$ scales with data span $T$ as
$T^{-5/2}$; moreover, data quality typically improves with time through
improved instrumentation. Also,
pulsar surveys continue to discover NS-WD systems
similar to B0655+64 but with shorter orbital periods and
millisecond pulse periods. Their higher rates of gravitational
energy release coupled with higher timing precision suggest that
these systems will surpass B0655+64 as laboratories for testing
theories of gravitation%.
---Table~\ref{nswd.tab} lists the current sample of relativistic NS-WD systems.
While B0655+64 remains a valuable object of further study, a potential
caveat applies to relativistic studies of all close NS-WD binaries:
very small orbital separations can give rise to
non-relativistic interactions. Heating of the companion star by
irradiation from the pulsar is thought to power tides and mass loss
in a handful of low-mass binaries. Outflows and tides cause
significant orbital torques (e.g., \citealp{aft94}), which would
overwhelm any small secular trend due to gravitational radiation. 
%\citet{vk95}
%suggest that such interactions may well be taking place
%even in the relatively-wide B0655+64 system, but in this case we
%know that orbital period changes have {\em not\/} been seen in the
%20 years since the system was discovered.
If indeed the closest NS-WD binaries (e.g., PSR B0751+18; see Nice
et al., this volume) are ``clean'' systems, significant new
constraints on SEP violation will emerge in the coming years. 

%The existing dataset shows evidence for arrival time delays at
%superior conjunction of the pulsar, of a few tens of microseconds.
%These can be modeled as a small eccentricity at the 3$\sigma$ level,
%but the implied orientation of the orbit (semi-major axis along the
%line of sight) is suggestive.  Some portion of the delay may be due
%to Shapiro delay in the companion's gravitational field, an
%important post-Keplerian measurable, rather than to a light
%travel-time difference in an elliptical orbit. Another possibility
%is that the delays are due to propagation through an ionized wind
%emitted by the companion star, if heating from the pulsar is
%important.  Future high-precision, multi-frequency timing can
%discriminate between these possibilities: delays due to propagation
%will be frequency-dependent, while Shapiro and light travel-time
%delays across a non-circular orbit will not. Similar timing of other
%close NS-WD systems will be needed to ensure that heated outflows
%are not contributing to any perceived orbital period variations.

\acknowledgments
I thank my collaborators, R. Dewey, A. G. Lyne, D. J. Nice, J. H.
Taylor, and S. E. Thorsett, for their continued interest and
participation in this project. The National Radio Astronomy
Observatory is a facility of the National Science Foundation
operated under cooperative agreement by Associated Universities,
Inc.

\bibliographystyle{apj}
\bibliography{apj-jour,myapj-jour,psrrefs,./0655} 
%\bibliography{apj-jour,journals,psrrefs,./0655} 

\end{document}